\documentclass[preprint,5p,twocolumn,10pt]{elsarticle}

\usepackage[pdftex,hypertexnames=false,linktocpage=true]{hyperref}
\hypersetup{pdfauthor={Lin Bolang},colorlinks=true,linkcolor=red,anchorcolor=darkgreen,citecolor=green!50!blue,filecolor=darkgreen,urlcolor=green,bookmarksnumbered=true,pdfview=FitB}
\usepackage{lipsum}
\usepackage{mathtools,cuted}
\usepackage{cuted}

\usepackage{graphicx}
\usepackage{float}
\usepackage{subfigure}

\usepackage{array}
\usepackage{booktabs} 
\usepackage{arydshln}
\usepackage{multirow}
\usepackage{makecell} 

\newcommand{\be}{\begin{equation}}  
\newcommand{\ee}{\end{equation}}  

\newcommand{\ba}{\begin{align}}  
\newcommand{\ea}{\end{align}}

\newcommand{\ra}{\rangle}
\newcommand{\la}{\langle}
\newcommand{\ma}{\Big{|}}
\newcommand{\nn}{\nonumber}

\hypersetup{colorlinks=true, citecolor=blue, urlcolor=blue, linkcolor=blue}

\usepackage{braket}

\usepackage{comment}
\begin{document}

\begin{frontmatter}
\title{Gluon skewed generalized parton distributions of proton from a light-front Hamiltonian approach}

\author[lzu,imp]{Pengxiang~Zhang}
\ead{zhangpx2021@lzu.edu.cn}

\author[imp,ucas]{Yiping~Liu}
\ead{liuyiping@impcas.ac.cn}

\author[imp,ucas,iowa]{Siqi~Xu}
\ead{xsq234@impcas.ac.cn}

\author[imp,ucas]{Chandan~Mondal\corref{cor1}}
\ead{mondal@impcas.ac.cn}

\author[imp,ucas]{Xingbo~Zhao}
\ead{xbzhao@impcas.ac.cn}

\author[iowa]{James~P.~Vary}
\ead{jvary@iastate.edu}

\author[]{\\\vspace{0.2cm}(BLFQ Collaboration)}

\address[lzu]{Lanzhou University, Lanzhou, Gansu, 730000, China}
\address[imp]{Institute of Modern Physics, Chinese Academy of Sciences, Lanzhou, Gansu, 730000, China}
\address[ucas]{School of Nuclear Physics, University of Chinese Academy of Sciences, Beijing, 100049, China}

\address[iowa]{Department of Physics and Astronomy, Iowa State University, Ames, IA 50011, USA}
\cortext[cor1]{Corresponding author}

\begin{abstract}
We calculate all leading-twist gluon generalized parton distributions (GPDs) inside the proton at nonzero skewness using the basis light-front quantization framework. The proton's light-front wave functions are derived from a light-front quantized Hamiltonian incorporating Quantum Chromodynamics inputs. Our results show that the qualitative behaviors of the GPDs are consistent with those from other theoretical calculations. Additionally, we analyze the GPDs in the boost-invariant longitudinal coordinate, $\sigma=\frac{1}{2} b^- P^+$, which serves as the Fourier conjugate of the skewness. The GPDs in $\sigma$-space exhibit diffraction patterns, reminiscent of optical wave diffraction.
\end{abstract}
\end{frontmatter}

\section{Introduction\label{Sec1}}
The proton, a bound state of quarks and gluon held together by the strong interaction described by quantum chromodynamics (QCD), is a central focus in nuclear and particle physics. Understanding its structure in terms of QCD's fundamental degrees of freedom is a major challenge. Over the past two decades, extensive experimental and theoretical efforts (see Ref.~\cite{Diehl:2023nmm} and the references therein) have been dedicated to studying generalized parton distributions (GPDs), which capture the three-dimensional (3D) spatial structure of the proton.

Precise knowledge of GPDs is essential for analyzing and interpreting exclusive scattering processes, such as deeply virtual Compton scattering (DVCS)~\cite{Ji:1996nm,Goeke:2001tz}, deeply virtual meson production (DVMP)~\cite{Goloskokov:2007nt,Collins:1996fb,Goloskokov:2024egn}, neutrino/electroweak meson production~\cite{Pire:2017yge,Pire:2021dad} single diffractive hard exclusive processes (SDHEPs)~\cite{Qiu:2022pla,Grocholski:2022rqj,Duplancic:2022ffo}, timelike Compton scattering (TCS)~\cite{Berger:2001xd}, wide-angle Compton scattering (WACS)~\cite{Radyushkin:1998rt,Diehl:1998kh}, and double DVCS~\cite{Deja:2023ahc}. Experimental collaborations worldwide, including Hall A~\cite{JeffersonLabHallA:2006prd, JeffersonLabHallA:2007jdm}, CLAS~\cite{CLAS:2001wjj, CLAS:2006krx, CLAS:2007clm} at Jefferson Lab (JLab), ZEUS~\cite{ZEUS:1998xpo, ZEUS:2003pwh}, H1~\cite{H1:2001nez, H1:2005gdw},  HERMES~\cite{HERMES:2001bob, HERMES:2006pre, HERMES:2008abz} at DESY, and COMPASS~\cite{dHose:2004usi} at CERN, etc., have made significant strides in investigating GPDs. 
GPDs provide crucial insights into the spatial distribution, spin, and orbital motion of partons inside the proton. Consequently, precise determination of proton GPDs is a primary objective for upcoming facilities like the Electron-Ion Colliders (EICs)~\cite{Accardi:2012qut,AbdulKhalek:2021gbh,Anderle:2021wcy,AbdulKhalek:2022hcn,Abir:2023fpo,Amoroso:2022eow,Hentschinski:2022xnd}, the Large Hadron-Electron Collider (LHeC)~\cite{LHeCStudyGroup:2012zhm,LHeC:2020van}, and the $12$ GeV upgrade at JLab~\cite{Dudek:2012vr, Burkert:2018nvj,Accardi:2023chb}.

 GPDs depend on the longitudinal momentum fraction ($x$) carried by the partons, the longitudinal momentum transfer ($\xi$) known as skewness, and the square of the total momentum transfer ($t$). In the forward limit, where $t=0$ and $\xi=0$, GPDs reduce to ordinary parton distribution functions (PDFs), which are accessible in deep inelastic scattering (DIS) experiments. The Mellin moments of GPDs correspond to form factors, and the Fourier transform of GPDs at zero skewness with respect to transverse momentum transfer yields impact parameter-dependent parton distributions, revealing how partons with specific longitudinal momentum are distributed in transverse position space~\cite{Burkardt:2000za,Burkardt:2002hr}. These distributions, unlike GPDs in momentum space, have a probabilistic interpretation and adhere to certain positivity conditions~\cite{Ralston:2001xs}. For nonzero skewness, GPDs can also be expressed in longitudinal position space via Fourier transformation with respect to the skewness variable $\xi$~\cite{Brodsky:2006in,Brodsky:2006ku,Chakrabarti:2008mw,Manohar:2010zm,Kumar:2015fta,Mondal:2015uha,Chakrabarti:2015ama,Mondal:2017wbf}.

From a theoretical standpoint, various nonperturbative techniques have been employed to explore quark GPDs, often taking a more phenomenological approach~\cite{Pasquini:2005dk,Pasquini:2006dv,Meissner:2009ww,Boffi:2002yy,Scopetta:2003et,Choi:2001fc,Choi:2002ic,Kaur:2023lun}. Promising methods for obtaining GPDs include Euclidean lattice QCD~\cite{Ji:2013dva,Ji:2020ect,Lin:2021brq,Lin:2020rxa,Bhattacharya:2022aob, Alexandrou:2021bbo, Alexandrou:2022dtc, Guo:2022upw, Alexandrou:2020zbe, Gockeler:2005cj, QCDSF:2006tkx, Alexandrou:2019ali,Hannaford-Gunn:2024aix}. While the study of gluon GPDs has not been as extensive as that of quark GPDs, recent research has increasingly focused on both chiral even and chiral odd gluon GPDs, particularly using light-cone spectator models~\cite{Tan:2023kbl,Chakrabarti:2024hwx,Chakrabarti:2023djs}, basis light-front quantization (BLFQ)~\cite{Lin:2023ezw,Lin:2024ijo}, and light-front holography~\cite{Gurjar:2022jkx,deTeramond:2021lxc}, double distribution representation~\cite{Goloskokov:2024egn}, and holographic string-based
approach~\cite{Mamo:2024jwp,Mamo:2024vjh} to explore these leading-twist phenomena.

In this work, we explore all the leading-twist gluon GPDs of the proton at nonzero skewness using the  nonperturbative framework, known as BLFQ~\cite{Vary:2009gt,Zhao:2014xaa,Li:2015zda,Nair:2022evk,Lan:2019vui,Mondal:2019jdg,Xu:2021wwj,Lan:2021wok,Xu:2023nqv}. While BLFQ has been previously applied to study gluon GPDs at zero skewness~\cite{Lin:2023ezw,Lin:2024ijo}, it is crucial to investigate GPDs at nonzero skewness, as experiments typically probe at $\xi \ne 0$. Within the BLFQ framework, an effective light-front Hamiltonian is used to determine the mass eigenstates and light-front wave functions (LFWFs), extending beyond the valence three-quark Fock sector ($\ket{qqq}$) to include the proton Fock sector with one dynamical gluon ($\ket{qqqg}$)\cite{Xu:2023nqv,Yu:2024mxo,Zhu:2024awq}. We consider the QCD interaction applicable to both Fock sectors~\cite{Brodsky:1997de}, supplemented by model confining potentials acting in both the transverse and longitudinal directions~\cite{Li:2015zda}. The resulting LFWFs have been successfully employed to study various proton properties, including electromagnetic form factors, radii, PDFs, GPDs, TMDs, and spin and orbital angular momentum~\cite{Xu:2023nqv,Yu:2024mxo,Zhu:2024awq}. In this study, we extend our investigations to examine gluon GPDs at nonzero skewness.

\section{Proton LFWFs with a dynamical gluon\label{Sec2}}
The LFWFs of the proton are obtained from the light-front\footnote{We adopt the light-front convention for the four-vector $v=\left( v^+, v^- , v^{\perp}\right)$, where $v^{\pm} = v^0 \pm v^3$ and $v^{\perp} = \left(v^1,v^2 \right)$.} Hamiltonian by solving the eigenvalue equation: 
$
H_{\rm LF}\vert \Psi\rangle=M^2\vert \Psi\rangle,
$
where $H_{\rm LF}=P^+ P^- - \vec{P}_\perp^{~2}$ represents the  Hamiltonian of the proton with $P^-$ being the light-front Hamiltonian, $P^+$ the longitudinal momentum, $\vec{P}_\perp$ the transverse momentum, and $M$ is the invariant mass of the system. At constant light-front time $x^+=x^0+x^3$, the proton state is expressed using various Fock sectors:
\begin{equation}
	\ket{\Psi}=\psi_{qqq}\ket{qqq}+\psi_{qqqg}\ket{qqqg}+\cdots,
	\label{fockeq}
\end{equation}
where $\psi_{(\cdots)}$ represents the LFWFs corresponding to each Fock state $\ket{\cdots}$. For practical computation, it is necessary to truncate the Fock space expansion to a finite number of dimensions.  At the model scale, we describe the proton through the LFWFs for the valence quarks, denoted as $\psi_{uud}$, as well as configurations incorporating three quarks and an additional dynamical gluon, represented by $\psi_{uudg}$. We employ a light-front Hamiltonian, $P^-=P^-_0+P^-_{\rm I}$, where $P^-_0$ represents the light-front QCD Hamiltonian corresponding to the $\ket{qqq}$ and $\ket{qqqg}$ Fock states of the proton, and $P^-_{\rm I}$ signifies a model Hamiltonian for the confining interaction potential~\cite{Xu:2023nqv}.

 In the light-front gauge $A^+ = 0$, the light-front QCD
Hamiltonian with a dynamical gluon is given by~\cite{Xu:2023nqv,Lan:2021wok},
\begin{equation}
	\begin{split}
		P_0^-= &\int \mathrm{d}x^- \mathrm{d}^2 x^{\perp} \Big\{\frac{1}{2}\bar{\psi}\gamma^+\frac{m_{0}^2+(i\partial^\perp)^2}{i\partial^+}\psi \\
		&+\frac{1}{2}A_a^i\left[m_g^2+(i\partial^\perp)^2\right] A^i_a +g_c\bar{\psi}\gamma_{\mu}T^aA_a^{\mu}\psi \\
		&+ \frac{1}{2}g_c^2\bar{\psi}\gamma^+T^a\psi\frac{1}{(i\partial^+)^2}\bar{\psi}\gamma^+T^a\psi \Big\}.
	\end{split}
	\label{hamieq}
\end{equation}
The first two terms in Eq.~(\ref{hamieq}) represent the kinetic energy of quarks and gluons, with bare masses $m_0$ for quarks and $m_g$ for gluons. Here, $\psi$ denotes the quark field, and $A^i_a$ represents the gluon field. The Dirac matrices are denoted by $\gamma^\mu$, and $T$ is the generator of the SU$(3)$ gauge group in color space. Although gluons are massless in standard QCD, our model introduces a phenomenological gluon mass to better capture low-energy phenomena as discussed in Ref.~\cite{Xu:2023nqv}. The third and fourth terms in Eq.~(\ref{hamieq}) describe quark-gluon interactions through vertex and instantaneous gluon interactions, governed by the coupling constant $g_c$. Following Fock-sector dependent renormalization~\cite{Perry:1990mz,Karmanov:2008br,Kurganov:2024tdn}, a mass counter-term ($\delta m_q$) is added to adjust the quark mass within the leading Fock sector to its renormalized value, $m_q = m_0 - \delta m_q$. Additionally, the model permits an independent quark mass $m_f$ in vertex interactions as suggested in Refs.~\cite{Glazek:1992aq,Burkardt:1998dd}.

The confining potential within the leading Fock sector, which includes both transverse and longitudinal components, is given by~\cite{Xu:2023nqv,Lan:2021wok},
\begin{align}
	P^-_IP^+=\frac{\kappa^4}{2}\sum\limits_{i\neq j}\left\{\Vec{r}_{ij\perp}^{\,2}-\frac{\partial_{x_i}(x_ix_j\partial_{x_j})}{(m_i+m_j)^2}\right\},
\end{align}
where $\Vec{r}_{ij\perp}$ quantifies the transverse distance between quarks, and $\kappa$ represents the strength of the confinement. For the $\ket{qqqg}$ Fock state, our model does not include an explicit confinement term. Instead, it relies on the effective mass of the gluon and the selected basis to account for confinement effects at the scale of the model.

In the BLFQ approach~\cite{Vary:2009gt}, we use two-dimensional harmonic oscillator (2D-HO) basis functions $\Phi_{nm}({p}_\perp;b)$ with the energy scale $b$, defined by radial ($n$) and angular ($m$) quantum numbers, to describe transverse degrees of freedom~\cite{Li:2015zda}. For longitudinal motion, plane-wave basis functions are employed within a box of length $2L$, with anti-periodic (fermions) or periodic (bosons) boundary conditions. Longitudinal momentum is given by $p^+ = \frac{\pi}{L}k$, where $k$ is a half-integer (integer) for fermions (bosons). We omit the zero mode for bosons. Each single-particle basis state is characterized by four quantum numbers $\ket{\alpha_i} = \ket{k_i, n_i, m_i, \lambda_i}$, where $\lambda$ denotes the light-front helicity. Multi-body basis states are constructed as the direct product of single-particle states, $\ket{\alpha} = \otimes_i \ket{\alpha_i}$ and we ensure a total color singlet configuration. All Fock-sector basis states share the same total angular momentum projection  $M_j$, satisfying the condition $\sum_i(\lambda_i+m_i)=M_j$.

We impose cutoffs, $\mathcal{K}$ and $\mathcal{N}_{\mathrm{max}}$, to constrain the basis in the longitudinal and transverse directions. The total longitudinal momentum is given by $\mathcal{K} = \sum_i k_i$, with each parton's momentum fraction $x_i = k_i / \mathcal{K}$. The transverse energy cutoff, $\mathcal{N}_{\mathrm{max}} \geq \sum_i (2n_i + |m_i| + 1)$, sets the range for 2D-HO states and determines the infrared (IR) and ultraviolet (UV) cutoffs, approximated as $\lambda_{\mathrm{IR}} \approx b / \sqrt{\mathcal{N}_{\mathrm{max}}}$ and $\lambda_{\mathrm{UV}} \approx b \sqrt{\mathcal{N}_{\mathrm{max}}}$, respectively~\cite{Zhao:2014xaa}. 

Diagonalizing the light-front Hamiltonian matrix provides the mass-squared spectra, $M^2$, and the LFWFs in momentum space:
\begin{equation}\label{wfs}
	\Psi^{M_J}_{N,\{x_i,\vec{p}_{i\perp},\lambda_i\}}=\sum_{\{n_i m_i\}}\psi^{M_J}_{N}(\{{\alpha}_i\})\prod_{i=1}^{N}  \Phi_{n_i m_i}(\vec{p}_{i\perp},b),
\end{equation}
where $\psi^{M_J}_{N=3}({\alpha_i})$ and $\psi^{M_J}_{N=4}({\alpha_i})$ correspond to the $\ket{uud}$ and $\ket{uudg}$ sectors, respectively.
In this paper, all calculations are performed with $N_{\rm max}=9$ and $K=16.5$. The effective Hamiltonian parameters are chosen to reproduce the proton mass and fit the flavor form factors~\cite{Xu:2023nqv}. The LFWFs, suitable for a low-resolution probe at $\mu_0^2\sim 0.24\pm 0.01$ GeV$^2$~\cite{Xu:2023nqv}, have been successfully applied to compute a wide range of proton observables, such as electromagnetic form factors, radii, PDFs, GPDs, TMDs, and spin and orbital angular momentum, etc.,~\cite{Xu:2023nqv,Yu:2024mxo,Zhu:2024awq}.

 \begin{table}[htp]
 \centering
 \caption{The nucleon's GPDs depending on the polarization of the nucleon (N) and the gluon (g) with $\bar{E}_T=2\widetilde{H}_T+E_T$. The symbols U, L, and T represent the nucleon polarizations, i.e., unpolarized, longitudinally polarized, and transversely polarized, respectively and U, Circular, and Linear represent the gluon polarizations, i.e., unpolarized, circularly polarized, and linearly polarized, respectively.}
 \vspace{0.15cm}
 \begin{tabular}{|  l |  c    c    c  |}
 \cline{1-4}
N/g  & U & Circular & Linear \\
 \cline{1-4} 
 U & $H$ & & $\bar{E}_T$ \\
 L &  & $\tilde{H} $ & $\tilde{E}_T$ \\
 T & E & $\tilde{E} $ & $H_T$, $\tilde{H}_T$ \\
 \cline{1-4} 
\end{tabular}
\label{table:GPDs}
\end{table}
\section{Gluon generalized parton distributions\label{Sec3}}
At leading twist, the proton has eight gluon GPDs: four chirally even ($H$, $E$, $\tilde{H}$, and $\tilde{E}$) and four chirally odd ($H_T$, $E_T$, $\tilde{H}_T$, and $\tilde{E}_T$). These GPDs relate to the polarization of the proton and gluons, as summarized in Table~\ref{table:GPDs}, and are defined through the off-forward matrix elements of the bilocal gluon tensor operator between proton states~\cite{Diehl:2003ny}:
\begin{align}
F^{g}_{\Lambda,\Lambda'} &= \frac{1}{\bar{P}^+}\int \frac{dz^-}{2\pi} e^{ix\bar{P}^+z^-} \nn \\
 &\times \la P',\Lambda' \ma F^{+i} \left(-\frac{z}{2} \right) F^{+i}  \left(\frac{z}{2} \right)  \ma P,\Lambda \ra \Big{|}_{\substack{z^+=0\\z^{\perp}=0}}  \label{unpol}\\
&=  \frac{1}{2\bar{P}^+} \bar{u}(P',\Lambda')\left[ H^g \gamma^+ +  E^g  \frac{i\sigma^{+\alpha} \left(-\Delta_{\alpha}\right)}{2M}\right] u(P,\Lambda),  \nn
\end{align}
\begin{align}
\tilde{F}^{g}_{\Lambda,\Lambda'} &= \frac{-i}{\bar{P}^+} \int \frac{dz^-}{2\pi} e^{ix\bar{P}^+z^-} \nn \\
&\times \la P',\Lambda' \ma F^{+i} \left(-\frac{z}{2} \right) \tilde{F}^{+i}  \left(\frac{z}{2} \right)  \ma P,\Lambda \ra \Big{|}_{\substack{z^+=0\\z^{\perp}=0}}  \label{helicity}\\
&=  \frac{1}{2\bar{P}^+}\bar{u}(P',\Lambda')\left[ \tilde{H}^g \gamma^+ \gamma_5 + \tilde{E}^g  \frac{\gamma_5 \left(-\Delta_{\alpha}\right)}{2M}\right] u(P,\Lambda),\nn
\end{align}
\begin{align}
F^{g}_{T\Lambda,\Lambda'} &= -\frac{1}{\bar{P}^+}\int \frac{{\rm d}z^-}{2\pi} e^{ix\bar{P}^+z^-}\nn \label{transversity}\\ 
&\times	\la P',\Lambda' | \mathbf{S} F^{+i} \left(-\frac{z}{2}\right) F^{j+} \left(\frac{z}{2} \right) | P,\Lambda \ra \Big{|}_{\substack{z^+=0\\z^{\perp}=0}} \nn \\
	&= \frac{\mathbf{S}}{2 \bar{P}^{+}} \frac{[\bar{P},\Delta]^{+j}}{2 M\bar{P}^+}\bar{u}( P^{\prime}, \Lambda^{\prime}) \left[i \sigma^{+i} H_T^g+\frac{[\gamma, \Delta]^{+i} }{2 M} E_T^g \right.\nn \\  
	&+ \left. \frac{[\bar{P}, \Delta]^{+i} }{M^2} \tilde{H}_T^g +\frac{[\gamma,\bar{P}]^{+i} }{M} \tilde{E}_T^g \right] u(P, \Lambda),
\end{align}
where $F^{+\mu}(x) = \partial^{+}A^{\mu}(x)$ corresponds to the gluon field tensor in the light-cone gauge and the dual field strength is given by $\tilde{F}^{\alpha \beta}(x) = \frac{1}{2}\epsilon^{\alpha \beta \gamma \delta }F_{\gamma \delta}(x)$. $\mathbf{S}$ stands for symmetrization of uncontracted Lorentz indices and removal of the trace. A summation over $i,\,j=1,\,2$ is implied. We use a compact notation $[a,b]^{+i} = a^+ b^i - b^+ a^i$.
In the symmetric frame, the average momentum of proton $\bar{P}= \frac{1}{2} (P^{\prime}+P)$, while momentum transfer $\Delta=(P^{\prime}-P)$. The initial and final four momenta of the proton are then given by
\begin{align}
P &\equiv \left((1+\xi)\bar{P}^+,\frac{M^2+\Delta_\perp^2/4}{(1+\xi)P^+},-\vec{\Delta}_\perp/2\right),\label{Pp}\\
P^{\prime} &\equiv \left((1-\xi)\bar{P}^+,\frac{M^2+\Delta_\perp^2/4}{(1-\xi)P^+},\vec{\Delta}_\perp/2\right). \label{Ppp}
\end{align}
Note that all the GPDs are functions of $x$, $\xi=- \Delta^+/2\bar{P}^+$, and $t= \Delta^2$. One can derive the following relation explicitly from $\Delta^-$
\begin{align}
- t= \frac{4 \xi^2 M^2 + \Delta_\perp^2}{(1-\xi^2)}\,. \label{mt_def}
\end{align}

In a reference frame where the momenta  $\vec{P}^{\prime}$ and $\vec{P}$ lie in the $x-z$ plane~\cite{Pasquini:2005dk}, we explicitly derive the relations for the chiral-even GPDs as follows:
\begin{equation}
    \begin{aligned}
        H^g(x, \xi, t)=&\frac{1}{\sqrt{1-\xi^2}} F^g_{++}+\frac{2 M \xi^2}{\sqrt{1-\xi^2} \Delta_{\perp 1}} F^g_{-+}, \\
        E^g(x, \xi, t)=&\frac{2 M \sqrt{1-\xi^2}}{\Delta_{\perp 1}}  F^g_{-+}, \\
       \tilde{H}^g(x, \xi, t)=&\frac{1}{\sqrt{1-\xi^2}} \widetilde{F}^g_{++}+\frac{2 M \xi}{\sqrt{1-\xi^2} \Delta_{\perp 1}} \widetilde{F}^g_{-+}, \\
       \tilde{E}^g(x, \xi, t)=&-\frac{2 M \sqrt{1-\xi^2}}{\xi \Delta_{\perp 1}} \widetilde{F}^g_{-+},\\
    \end{aligned}
\end{equation}
where the proton helicity  is denoted by $\Lambda=+(-)$, corresponding to $+1(-1)$, respectively. For the chiral-odd GPDs, we transform the matrix elements from the helicity basis to the transversity basis~\cite{Pasquini:2005dk,Chakrabarti:2015ama} and decompose the gluon tensor operator into $\theta_1=F^{+1}F^{1+}-F^{+2}F^{2+}$ and $\theta_2=i(F^{+1}F^{2+}+F^{+2}F^{1+})$~\cite{Diehl:2003ny}. In the transversity basis, we obtain 
\begin{align}
H_T^g(x,\xi,t)=&-\frac{2M}{\sqrt{1-\xi^2}\Delta_{\perp 1}}F^{g[\theta_1]}_{\uparrow\uparrow}-\frac{4M^2\xi}{\sqrt{1-\xi^2}\Delta_{\perp 1}^2}F^{g[\theta_1]}_{\downarrow\uparrow},\nn\\
E_T^g(x,\xi,t)=&\frac{4M^2}{\sqrt{1-\xi^2}\Delta_{\perp 1}^2}\Big(\xi F^{g[\theta_1]}_{\downarrow\uparrow}+F^{g[\theta_2]}_{\uparrow\uparrow}\Big)\nn\\&+\frac{8M^3}{\sqrt{1-\xi^2}\Delta_{\perp 1}^3}\Big(F^{g[\theta_2]}_{\uparrow\downarrow}-F^{g[\theta_1]}_{\uparrow\uparrow}\Big),\nn\\
\tilde{H}_T^g(x,\xi,t)=&-\frac{4M^3\sqrt{1-\xi^2}}{\Delta_{\perp 1}^3}\Big(F^{g[\theta_2]}_{\uparrow\downarrow}-F^{g[\theta_1]}_{\uparrow\uparrow}\Big),\\
\tilde{E}_T^g(x,\xi,t)=&\frac{4M^2}{\sqrt{1-\xi^2}\Delta_{\perp 1}^2}\Big(F^{g[\theta_1]}_{\downarrow\uparrow}+\xi F^{g[\theta_2]}_{\uparrow\uparrow}\Big)\nn\\&+\frac{8M^3\xi}{\sqrt{1-\xi^2}\Delta_{\perp 1}^3}\Big(F^{g[\theta_2]}_{\uparrow\downarrow}-F^{g[\theta_1]}_{\uparrow\uparrow}\Big),\nn
\end{align}
where
\begin{align}
F_{\Lambda_T^\prime, \Lambda_T}^{g[\theta_i]} = &-\frac{1}{\bar{P}^+}\int \frac{dz^-}{2\pi} e^{ix\bar{P}^+z^-} \nn\\&\times \la P',\Lambda_T'| \theta_i | P,\Lambda_T \ra \Big{|}_{\substack{z^+=0\\z^{\perp}=0}},
\label{helicity_amp}
\end{align}
with $\Lambda_T=\uparrow (\downarrow)$ being the transverse polarization of the proton polarized along the +ve $\hat{x}$ ($\uparrow$) or
-ve $\hat{x}$ ($\downarrow$) direction. In overlap representation, $F^{g}_{\Lambda\Lambda^\prime}$, $\widetilde{F}^{g}_{\Lambda\Lambda^\prime}$ and $F^{g[\theta_i]}_{\Lambda_T\Lambda^\prime_T}$  are expressed in terms of the LFWFs as
\begin{equation}
\begin{aligned}
F^{g}_{\Lambda^\prime,\Lambda}=&\sum_{\{\lambda_i\}} \int \left[{\rm d}\mathcal{X} \,{\rm d}\mathcal{P}_\perp\right]\, \Psi^{\Lambda^\prime *}_{4,\{x^{\prime\prime}_i,\vec{k}^{\prime\prime}_{i\perp},\lambda_i\}}\Psi^{\Lambda}_{4,\{x_i^{\prime},\vec{k}_{i\perp}^{\prime},\lambda_i\}},
\\
\widetilde{F}^{g}_{\Lambda^\prime,\Lambda}=& \sum_{\{\lambda_i\}} \int \left[{\rm d}\mathcal{X} \,{\rm d}\mathcal{P}_\perp\right]\, \lambda_1\,\Psi^{\Lambda^\prime *}_{4,\{x^{\prime\prime}_i,\vec{k}^{\prime\prime}_{i\perp},\lambda_i\}}\Psi^{\Lambda}_{4,\{x_i^{\prime},\vec{k}_{i\perp}^{\prime},\lambda_i\}},
\end{aligned}
\end{equation}
\begin{equation}
\begin{aligned}
F^{g[\theta_1]}_{\uparrow\uparrow}=& \sum_{\{\lambda_i,\,\lambda_i^\prime\}} \int \left[{\rm d}\mathcal{X} \,{\rm d}\mathcal{P}_\perp\right]\, \Psi^{- *}_{4,\{x^{\prime\prime}_i,\vec{k}^{\prime\prime}_{i\perp},\lambda_i^\prime\}}\Psi^{+}_{4,\{x_i^{\prime},\vec{k}_{i\perp}^{\prime},\lambda_i\}} \\&\times\delta_{\lambda_1^\prime,-\lambda_1}\delta_{\lambda_{2,3,4}^\prime,\lambda_{2,3,4}},
\\
F^{g[\theta_1]}_{\downarrow\uparrow}=& \sum_{\{\lambda_i,\,\lambda_i^\prime\}} \int \left[{\rm d}\mathcal{X} \,{\rm d}\mathcal{P}_\perp\right]\, \Psi^{+ *}_{4,\{x^{\prime\prime}_i,\vec{k}^{\prime\prime}_{i\perp},\lambda_i^\prime\}}\Psi^{+}_{4,\{x_i^{\prime},\vec{k}_{i\perp}^{\prime},\lambda_i\}} \\&\times \lambda_1^\prime \,\delta_{\lambda_1^\prime,-\lambda_1}\delta_{\lambda_{2,3,4}^\prime,\lambda_{2,3,4}},
\\
F^{g[\theta_2]}_{\uparrow\uparrow}=& \sum_{\{\lambda_i,\,\lambda_i^\prime\}} \int \left[{\rm d}\mathcal{X} \,{\rm d}\mathcal{P}_\perp\right]\, \Psi^{+ *}_{4,\{x^{\prime\prime}_i,\vec{k}^{\prime\prime}_{i\perp},\lambda_i^\prime\}}\Psi^{+}_{4,\{x_i^{\prime},\vec{k}_{i\perp}^{\prime},\lambda_i\}} \\&\times \delta_{\lambda_1^\prime,-\lambda_1}\delta_{\lambda_{2,3,4}^\prime,\lambda_{2,3,4}},
\\
F^{g[\theta_2]}_{\downarrow\uparrow}=& \sum_{\{\lambda_i,\,\lambda_i^\prime\}} \int \left[{\rm d}\mathcal{X} \,{\rm d}\mathcal{P}_\perp\right]\, \Psi^{- *}_{4,\{x^{\prime\prime}_i,\vec{k}^{\prime\prime}_{i\perp},\lambda_i^\prime\}}\Psi^{+}_{4,\{x_i^{\prime},\vec{k}_{i\perp}^{\prime},\lambda_i\}} \\&\times (-\lambda_1^\prime) \,\delta_{\lambda_1^\prime,-\lambda_1}\delta_{\lambda_{2,3,4}^\prime,\lambda_{2,3,4}},
\end{aligned}
\end{equation}
where 
\begin{align}
\left[{\rm d}\mathcal{X} \,{\rm d}\mathcal{P}_\perp\right]_n=&(\sqrt{1-\xi^2}\ )^{2-n}\prod_{i=1}^n \left[\frac{{\rm d}x_i{\rm d}^2 \vec{k}_{i\perp}} {16\pi^3}\right]\delta(x-x_1)\nonumber\\
&\times16 \pi^3 \delta \left(1-\sum_{i=1}^{3} x_i\right) \delta^2 \left(\sum_{i=1}^{3}\vec{k}_{i\perp}\right),  
\end{align} 
and the light-front momenta are $x^{\prime}_1=\frac{x_1+\xi}{1+\xi}$; $\vec{k}^{\prime}_{1\perp}=\vec{k}_{1\perp}+(1-x^{\prime})\frac{\vec{\Delta}_{\perp}}{2}$ for the initial gluon ($i=1$) and $x^{\prime}_i=\frac{x_i}{1+\xi}; ~\vec{k}^{\prime}_{i\perp}=\vec{k}_{i\perp}-{x_i^{\prime}} \frac{\vec{\Delta}_{\perp}}{2}$ for the initial spectators  ($i\ne1$) and
$x^{\prime\prime}_1=\frac{x_1-\xi}{1-\xi}$; $\vec{k}^{\prime}_{1\perp}=\vec{k}_{1\perp}-(1-x^{\prime\prime})\frac{\vec{\Delta}_{\perp}}{2}$ for the final gluon and $x^{\prime\prime}_i=\frac{x_i}{1-\xi}; ~\vec{k}^{\prime}_{i\perp}=\vec{k}_{i\perp}+{x_i^{\prime\prime}} \frac{\vec{\Delta}_{\perp}}{2}$ for the final spectators and
$\lambda_1 (\lambda_1^\prime)$ designates the gluon helicity.
Note that, in this work, we limit ourselves to the Dokshitzer-Gribov-Lipatov-Altarelli-Parisi (DGLAP) region, $\xi<x<1$, where the number of partons in the initial and the final states remains conserved.

\begin{figure*}[htp]
\centering
\includegraphics[scale=.47]{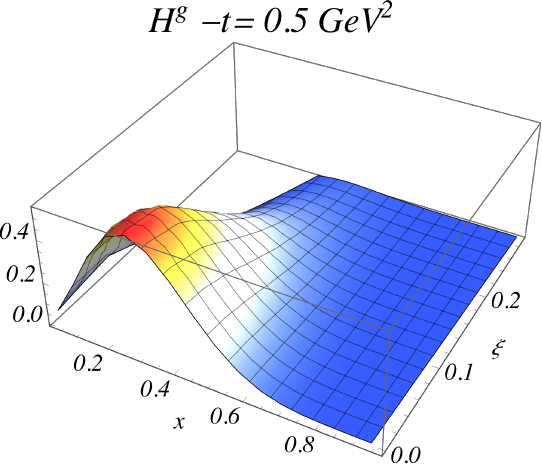}
\includegraphics[scale=.47]{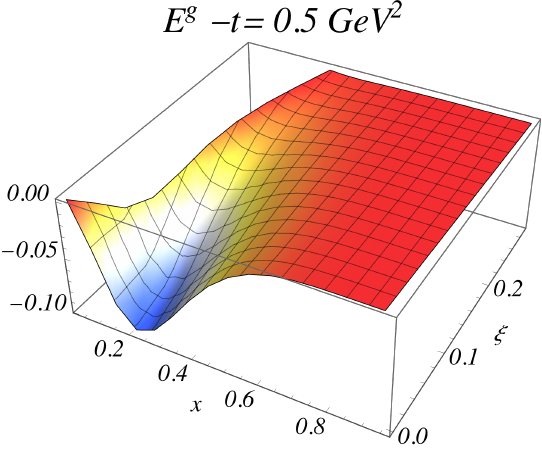} 
\includegraphics[scale=.47]{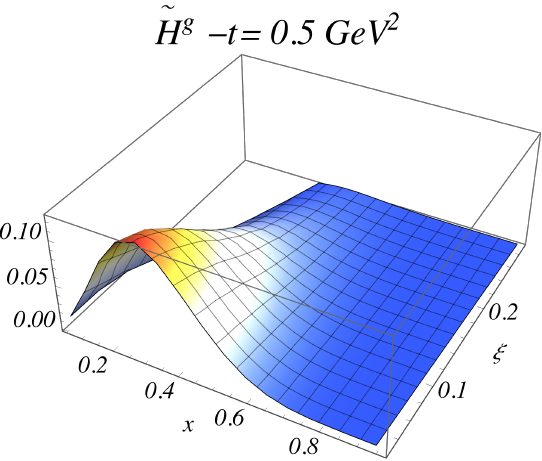} 
\includegraphics[scale=.47]{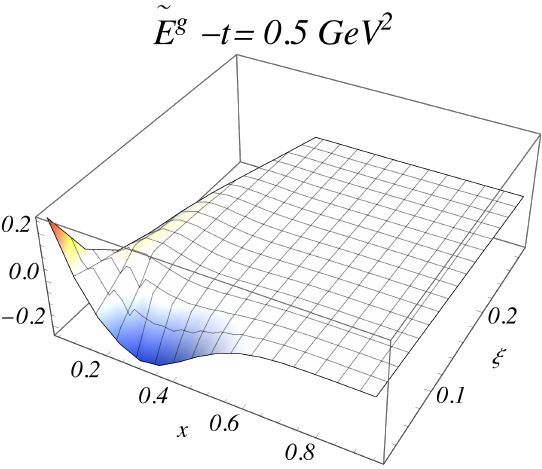} \\
\includegraphics[scale=.47]{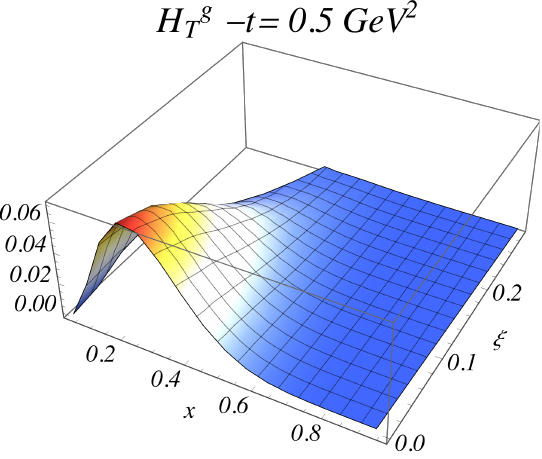}
\includegraphics[scale=.47]{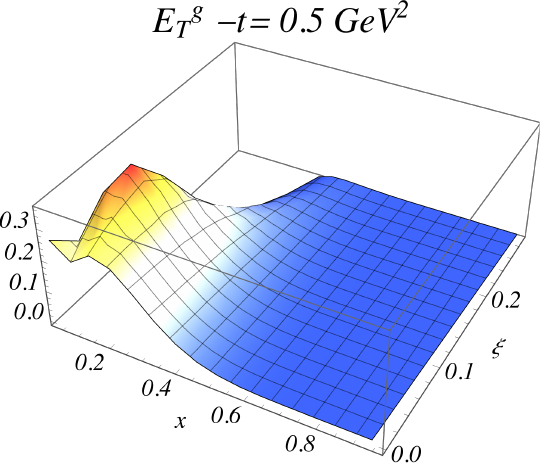} 
\includegraphics[scale=.47]{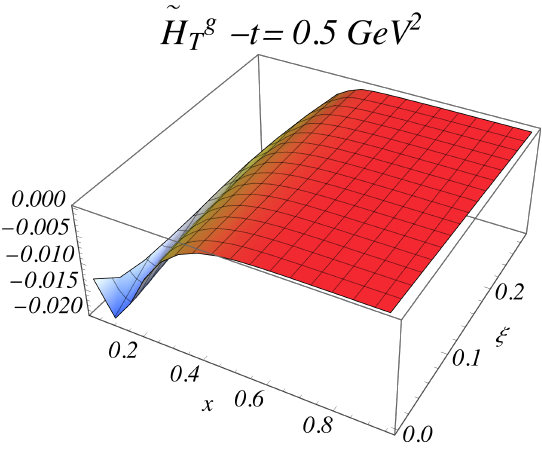} 
\includegraphics[scale=.47]{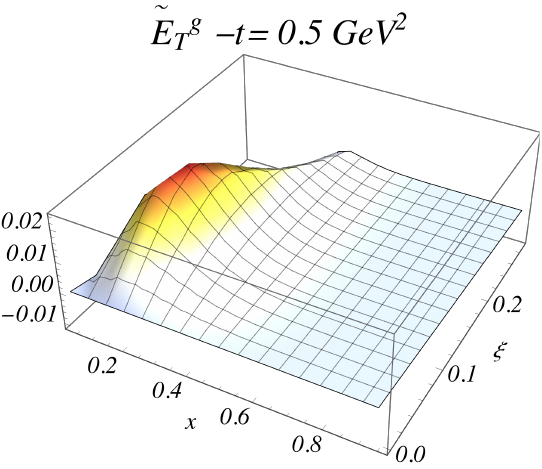} 
\caption{\label{chiral_even_xz} The gluon GPDs as functions of $x$ and  $\xi$ for fixed $t=-0.5$ GeV$^2$. The upper (lower) panel is for the chiral-even (odd) GPDs.}
\end{figure*}
\section{Numerical results and discussions}
\label{rsult}

In our BLFQ approach, the use of a discretized plane-wave basis in the longitudinal direction results in discretized momentum fractions for the partons, \(x_i = \frac{k_i}{\mathcal{K}}\), where \(k_i\) are half-integers for the quarks and integers for the gluon. Calculating GPDs at nonzero skewness involves overlapping LFWFs with different longitudinal momenta. We interpolate the longitudinal component of our LFWFs to obtain all leading-twist GPDs. These GPDs are related to physical quantities such as orbital angular momentum, axial and tensor charges, and are also connected to form factors and PDFs in specific kinematic limits~\cite{Diehl:2003ny}. In presenting our numerical results, we focus on the \(\xi\)-dependence of the gluon GPDs, as the other dependencies with vanishing skewness have been explored in previous studies~\cite{Lin:2023ezw,Lin:2024ijo}.

\subsection{GPDs in momentum space} \label{sec_GPDs_momentum}

In Fig.~\ref{chiral_even_xz} (upper panel), we present the gluon chiral-even GPDs as functions of $x$ and $\xi$ for a fixed value of $t=-0.5$ GeV$^2$. The general features of these GPDs are similar, differing mainly by overall sign and scale, with the exception of $\tilde{E}^g$ (discussed below). All GPDs peak at lower $x \,(<0.5)$, with peaks shifting to higher $x$ and diminishing as the longitudinal momentum transfer increases. In the large $x$ region, all GPDs eventually decay and become independent of $\xi$. Both $H^g$ and $\tilde{H}^g$ show positive peaks along $\xi$, while $E^g$ shows negative peaks, with maximum values at $\xi=0$. The peak of $H^g$ is notably larger than those of $E^g$ and $\tilde{H}^g$, which are comparable in magnitude. The GPD $\tilde{E}^g$ behaves differently, crossing zero along $x$ at small $\xi$. The forward limits of $H^g$ and $\tilde{H}^g$ correspond to spin-independent and spin-dependent gluon PDFs, respectively, as reported in Ref.~\cite{Xu:2023nqv}. In contrast, $E^g$ and $\tilde{E}^g$ have no forward limits and decouple in this regime~\cite{Diehl:2003ny}.

When comparing our results with those from the light-cone spectator models~\cite{Gurjar:2022jkx,Tan:2023kbl}, a significant difference emerges. In our case, both $H^g$ and $\widetilde{H}^g$ are positive definite, whereas the spectator models exhibit negative regions at small $x$ and oscillatory behavior in $x$ space. However, the qualitative behavior of $E^g$ and $\tilde{E}^g$ in Ref.~\cite{Gurjar:2022jkx} is consistent with our findings. Additionally, the overall behavior of our gluon GPDs is similar to those of quark GPDs in the proton~\cite{Liu:2024umn,Ji:1997gm, Boffi:2002yy, Boffi:2003yj, Mondal:2015uha, Mondal:2017wbf,Freese:2020mcx} and in the pion~\cite{Kaur:2018ewq}. Notably, within our BLFQ approach, the gluon GPDs decrease more rapidly with increasing $\xi$ compared to the quark GPDs~\cite{Liu:2024umn}.

\begin{figure*}[htp]
\centering
\includegraphics[scale=.47]{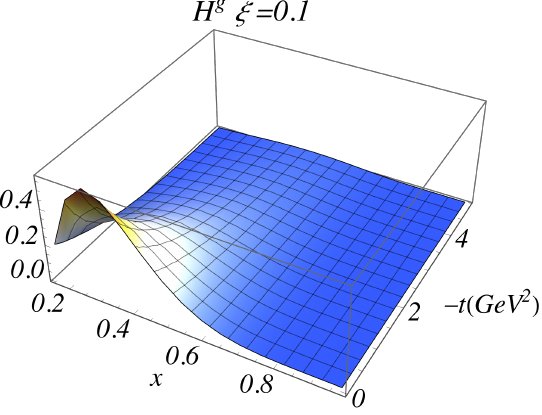}
\includegraphics[scale=.47]{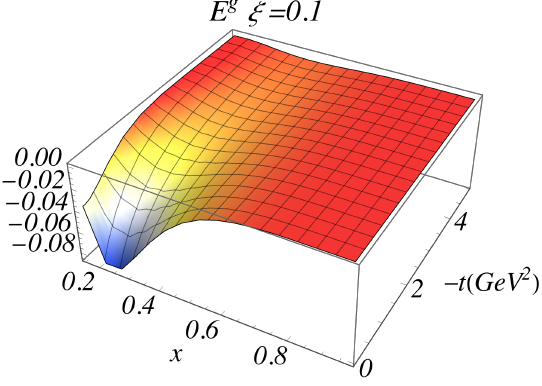}
\includegraphics[scale=.47]{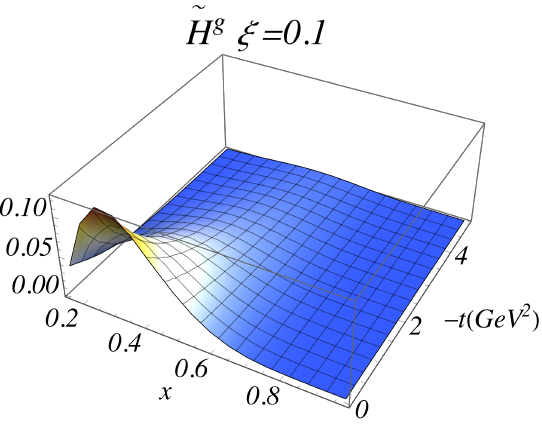} 
\includegraphics[scale=.47]{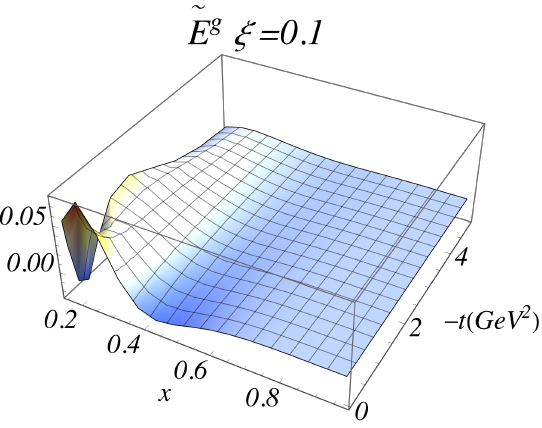} \\
\includegraphics[scale=.47]{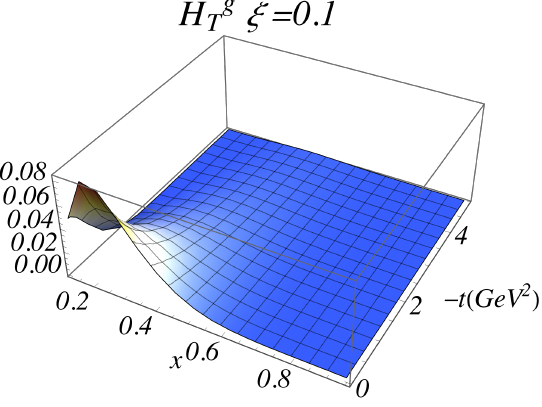}
\includegraphics[scale=.47]{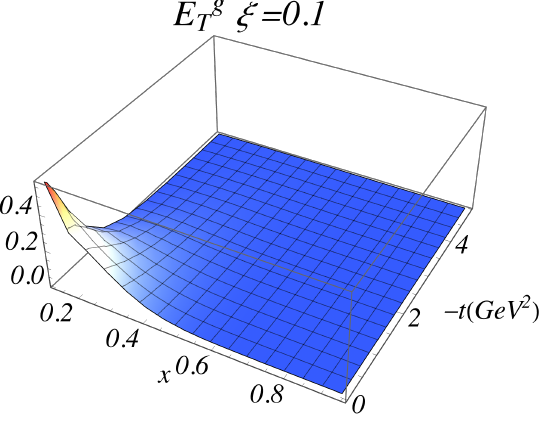}
\includegraphics[scale=.47]{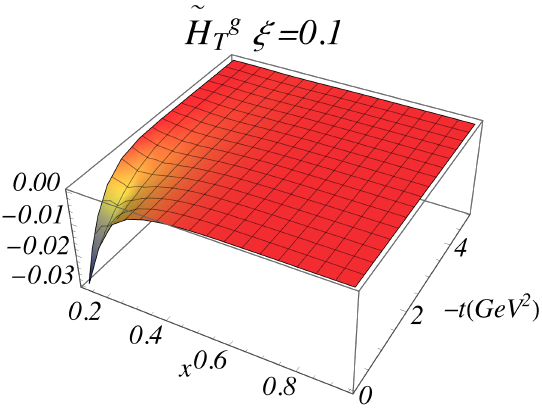} 
\includegraphics[scale=.47]{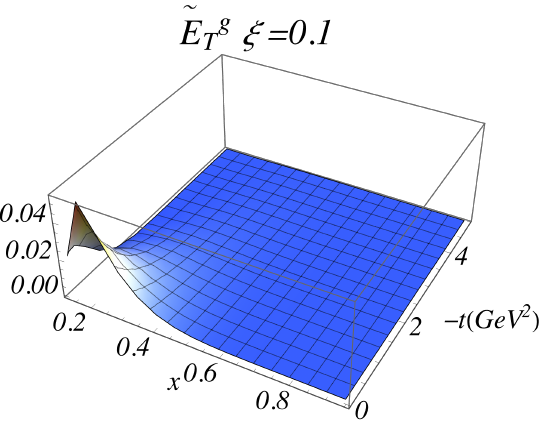} 
\caption{\label{chiral_even_xt} The gluon GPDs as functions of $x$ and $-t$ for fixed $\xi=0.1$. The upper (lower) panel is for the chiral-even (odd) GPDs.}
\end{figure*}

All the chiral-odd GPDs, shown in Fig.~\ref{chiral_even_xz} (lower panel), display similar qualitative behavior to the chiral-even GPDs, except for $\tilde{E}_T$. The GPD $\tilde{E}_T$ vanishes at $\xi=0$. It is an odd function under the transformation $\xi \rightarrow -\xi$~\cite{Diehl:2003ny}, satisfying $\tilde{E}_T(x,\,-\xi,\,t)=-\tilde{E}_T(x,\,\xi,\,t)$. Except for $\tilde{H}_T$, all other chiral-odd GPDs show positive distributions. ${H}^g_T$ and $\tilde{H}^g_T$ peak at $\xi=0$, while ${E}^g_T$ and $\tilde{E}^g_T$ have positive peaks along $\xi$, with extrema around $\xi\sim 0.1$ for $-t=0.5$ GeV$^2$. The peak shifts to larger $x$ with increasing $\xi$. We observe that the signs and qualitative features of $H_T^g$ and $E_T^g$ in our BLFQ analysis are consistent with those obtained using the quark target model~\cite{Meissner:2007rx} and light-cone spectator model~\cite{Chakrabarti:2024hwx}. Additionally, in our study, $\widetilde{H}_T^g$ is distinctly non-zero, differing from those phenomenological model outcomes. A notable distribution emerges from combining two chiral-odd GPDs, $2\tilde{H}^g_T + E^g_T$, which offers insights into transverse spin and angular momentum contributions in specific limits~\cite{Diehl:2005jf,Burkardt:2005hp,Lin:2024ijo}. 
All distributions show accessibility of the DGLAP region $x>\xi$, with the distribution width in $x$ decreasing as $\xi$ increases. The qualitative behavior of our gluon chiral-odd GPDs aligns with quark GPDs~\cite{Liu:2024umn,Pasquini:2005dk, Chakrabarti:2015ama}.

Figure~\ref{chiral_even_xt} shows the gluon chiral-even (upper panel) and chiral-odd  GPDs (lower panel), respectively, as functions of $x$ and $-t$ for a fixed $\xi=0.1$. The distributions peak when there is no transverse momentum transfer to the proton's final state and when the gluon carries less than 50\% of the proton's longitudinal momentum. As momentum transfer increases, the peak shifts towards higher $x$ values, while the magnitude decreases. In the large $x$ region, all distributions eventually decay and become independent of $-t$, with a faster decay for the chiral-odd GPDs compared to the chiral-even GPDs. We also notice that $E$ and $E_T$ fall faster than $H$ and $H_T$. We observe a bump in $\tilde{E}$ at $-t \sim 2 \, \mathrm{GeV}^2$, occurring at a value of $x$ where other GPDs nearly vanish. Notably, $\tilde{H}_T$ has a sharp negative peak at $x\sim\xi$ and decays more rapidly with $-t$ than $\tilde{E}_T$. These characteristics are model and parton-independent, consistent with observations in quark distributions in various QCD-inspired models~\cite{Ji:1997gm, Boffi:2002yy, Boffi:2003yj, Mondal:2015uha, Mondal:2017wbf, Freese:2020mcx, Pasquini:2005dk, Chakrabarti:2015ama} as well as in our BLFQ approach~\cite{Liu:2024umn}.

In the forward limit, GPDs reduce to PDFs, such as the unpolarized and helicity-dependent PDFs, with $H(x,0,0)=f_1(x)$ and $\tilde{H}(x,0,0)=g_1(x)$, respectively. Unlike the quark transversity PDF, $H_T(x,0,0)=h_1(x)$, a gluon transversity PDF does not exist for a spin-$\frac{1}{2}$ target, as gluon helicity flip requires targets with spin 1 or higher for angular momentum conservation. Studies on gluon PDFs, angular momentum, and generalized form factors corresponding to GPDs $H$, $E$, $\tilde{H}$, and $2\tilde{H}^q_T+E^q_T$ within our BLFQ approach are reported in Refs.~\cite{Xu:2023nqv,Lin:2023ezw,Lin:2024ijo}.

\begin{figure*}[htp]
\includegraphics[scale=.47]{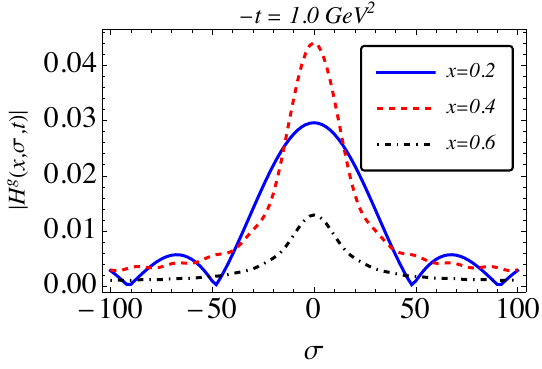}
\includegraphics[scale=.47]{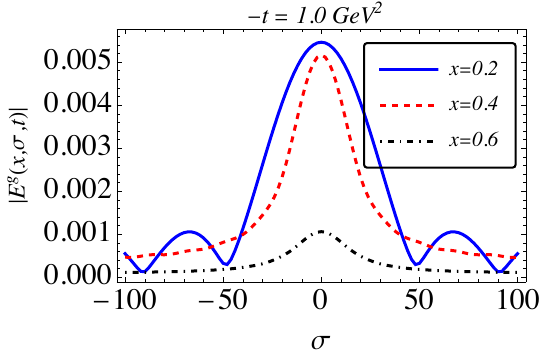}
\includegraphics[scale=.47]{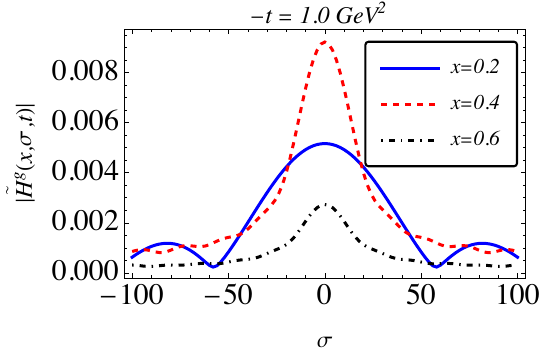}
\includegraphics[scale=.47]{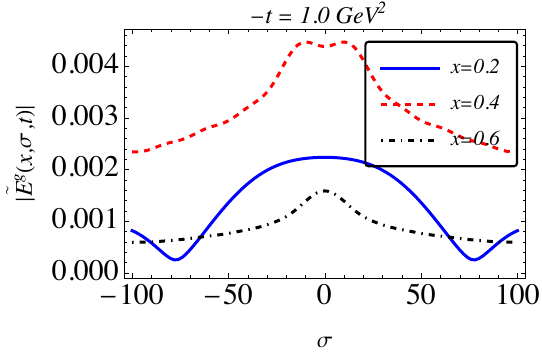} \\
\includegraphics[scale=.47]{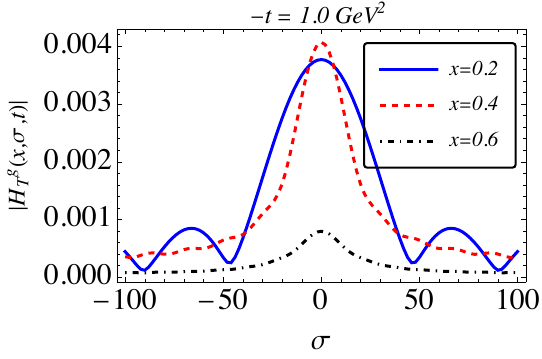}
\includegraphics[scale=.47]{FTabs_HTg_t_1p0_AllSector.pdf}
\includegraphics[scale=.47]{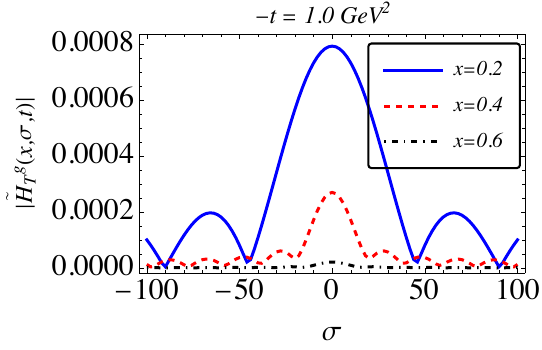}
\includegraphics[scale=.47]{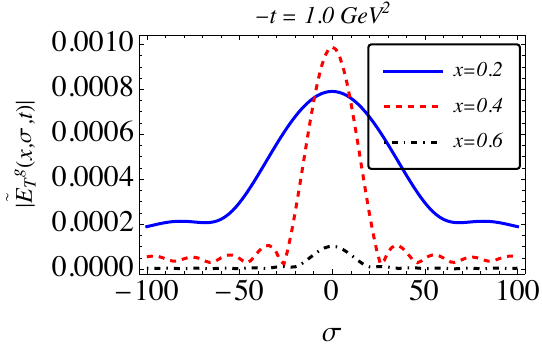}
\caption{\label{chiral_even_xs} The gluon GPDs in boost-invariant longitudinal position space as functions of $\sigma$ for  $x=\{0.2,\,0.4,\, 0.6\}$ and fixed $t=-1.0$ GeV$^2$. The upper (lower) panel is for the chiral-even (odd) GPDs.}
\end{figure*}

\subsection{GPDs in boost-invariant longitudinal space} \label{sec_GPDs_sigma}
The GPDs in transverse position space have been widely studied using various theoretical approaches, including the BLFQ framework at zero skewness. The transverse impact parameter $\vec{b}_\perp$ is the Fourier conjugate to the variable $\vec{D}_\perp=\vec{\Delta}_\perp/(1-\xi^2)$~\cite{Diehl:2002he,Burkardt:2002hr,Ralston:2001xs,Kaur:2018ewq}, which simplifies to $\vec{\Delta}_\perp$ at zero skewness.
Similarly, the longitudinal momentum transfer $\xi P^+$ is the Fourier conjugate to the longitudinal distance $\frac{1}{2}b^-$. Thus, $\xi$ is conjugate to the boost-invariant longitudinal impact parameter $\sigma=\frac{1}{2}b^-P^+$. Fourier transforming the GPDs with respect to $\xi$ provides distributions in boost-invariant longitudinal space $\sigma$, where the DVCS amplitude reveals a diffraction pattern~\cite{Brodsky:2006in,Brodsky:2006ku}, akin to optical diffraction. This makes exploring GPDs in longitudinal impact parameter space particularly intriguing. The GPDs in this space are defined as follows:
\begin{equation}
\begin{aligned}
f(x,\sigma,t)&=\int_0^{\xi_f}\frac{d\xi}{2\pi}e^{i\xi P^+b^-/2}G(x,\xi,t)\\
&=\int_0^{\xi_f}\frac{d\xi}{2\pi}e^{i\xi\sigma}G(x,\xi,t),
\end{aligned}
\label{LFT}
\end{equation}
where $G$ stands for the GPDs. 
The upper integration limit, $\xi_f$, plays a role analogous to the slit width in diffraction, setting a key condition for the emergence of the diffraction pattern. Since we focus on the region $\xi < x < 1$, the upper limit is defined as $\xi_f = \mathrm{min}\{x, \xi_{\mathrm{max}}\}$. For a fixed value of $-t$, the maximum value of $\xi$ is given by~\cite{Brodsky:2000xy}
\be
\xi_{\mathrm{max}} = \frac{1}{\left[1 + \frac{4M^2}{(-t)}\right]^{\frac{1}{2}}}.
\ee

Figure~\ref{chiral_even_xs} presents our results for the modulus of the gluon's chiral-even (upper panel) and chiral-odd GPDs (lower panel), respectively in longitudinal position space for three different values of $x = \{0.2,\,0.4,\,0.6\}$ with a fixed $t = -1.0$ GeV$^2$. The chosen value of $-t$ corresponds to $\xi_{\mathrm{max}} \approx 0.47$, setting the upper limit for the $\xi$ integration in Eq.~(\ref{LFT}). Specifically, $\xi_f = 0.2$ and $0.4$ for $x = 0.2$ and $0.4$, respectively, while for $x = 0.6$, the integration limit is $\xi_f = 0.47$ since $x > \xi_{\mathrm{max}}$ in this case.

Our results show oscillatory behavior similar to the diffraction pattern seen in a single-slit optical experiment, where the width of the principal maxima is inversely proportional to the slit width. In the Fourier transform described by Eq.~\eqref{LFT}, the finite range of $\xi$ generates this diffraction pattern, with $\xi_f$ acting as the slit width. However, not all functions with a finite range of $\xi$ produce such a pattern; for example, $\tilde{E}^g(x,\sigma,t)$ and $\tilde{E}^g_T(x,\sigma,t)$ behave differently compared to the other GPDs.

A major feature appears as $\xi_f$ increases, where the principal maxima narrow, with the first minima shifting inward. Similar patterns have been observed in DVCS amplitudes, coordinate-space parton densities, and Wigner distributions in various models~\cite{Brodsky:2006in,Brodsky:2006ku,Miller:2019ysh,Chakrabarti:2008mw,Manohar:2010zm,Kumar:2015fta,Mondal:2015uha,Chakrabarti:2015ama,Mondal:2017wbf,Kaur:2018ewq,Maji:2022tog,Ojha:2022fls}. Notably, gluon GPDs in longitudinal position space show different characteristics compared to quark GPDs. The diffraction pattern is prominent at small $x$ for gluons but transitions into a single peak as $x$ increases, while quark distributions retain the pattern even at large $x$. This difference arises from the more rapid decline of gluon GPDs with increasing $\xi$ compared to quark GPDs~\cite{Liu:2024umn}.  Additionally, all distributions in boost-invariant longitudinal space show a long-distance tail, as noted in Refs.~\cite{Miller:2019ysh,Weller:2021wog}.  We observe that gluon distributions exhibit longer tails compared to quark distributions~\cite{Liu:2024umn}.

\section{Conclusions}
The study of gluon skewness-dependent GPDs is of interest due to their connection with exclusive scattering cross sections, measurable in EIC and EicC experiments. Using BLFQ, a nonperturbative tool for solving many-body bound state problems in quantum field theory, we calculated all leading-twist, gluon skewed GPDs for the proton from its LFWFs. These wave functions were obtained from the eigenvectors of an effective QCD Hamiltonian, incorporating three-dimensional confinement and fundamental QCD interactions. We focused on the DGLAP region ($x>\xi$), where our BLFQ approach shows qualitative similarities with other phenomenological models. Interestingly, $\tilde{H}_T^g$ is non-zero, unlike other models~\cite{Meissner:2007rx,Chakrabarti:2024hwx,Tan:2023kbl}, and gluon GPDs decrease more rapidly with increasing $\xi$ compared to quark GPDs~\cite{Liu:2024umn}.

The gluon GPDs in longitudinal impact-parameter space offer a unique view of the proton's structure. By performing a Fourier transform of the skewed GPDs with respect to $\xi$, we obtained the GPDs in boost-invariant longitudinal position space, $\sigma = \frac{1}{2}b^-P^+$. These gluon GPDs exhibit a diffraction pattern similar to that observed in single-slit optical experiments, where the central maxima's width is inversely proportional to the slit width, represented here by the finiteness of $\xi_f$. However, this pattern is influenced by the functional behavior of the GPDs as well. Notably, gluon GPDs in longitudinal position space differ from quark GPDs, showing a prominent diffraction pattern at small $x$ that transitions into a single peak as $x$ increases, whereas quark distributions maintain the pattern even at large $x$. Additionally, gluon distributions exhibit longer tails compared to quark distributions~\cite{Liu:2024umn}.
A similar diffraction pattern has been observed in various observables, including the DVCS amplitude, parton density, and Wigner distributions in longitudinal position space.

We will extend our approach to include the $|qqqq\bar{q}\rangle$ and $|qqqgg\rangle$ Fock sectors, enabling the evaluation of skewed GPDs in the ERBL region ($x<\xi$) and the study of sea-quark GPDs. This approach will also be applied to calculate higher-twist GPDs~\cite{Zhang:2023xfe}, with future work focusing on genuine contributions beyond the Wandzura-Wilczek approximation.

\section*{Acknowledgements\label{Sec6}}
We thank Bolang Lin, Sreeraj Nair, Ziqi Zhang, Zhimin Zhu, and Zhi Hu for helpful discussions in the Institute of Modern Physics, University of Chinese Academy of Science. C. M. is also supported by new faculty start up funding by the Institute of Modern Physics, Chinese Academy of Sciences, Grant No. E129952YR0. X. Zhao is supported by new faculty startup funding by the Institute of Modern Physics, Chinese Academy of Sciences, by Key Re- search Program of Frontier Sciences, Chinese Academy of Sciences, Grant No. ZDBS-LY-7020, by the Foundation for Key Talents of Gansu Province, by the Central Funds Guiding the Local Science and Technology Development of Gansu Province, Grant No. 22ZY1QA006, by international partnership program of the Chinese Academy of Sciences, Grant No. 016GJHZ2022103FN, by National Key R\&D Program of China, Grant No. 2023YFA1606903, by the Strategic Priority Research Program of the Chinese Academy of Sciences, Grant No. XDB34000000, and by the National Natural Science Foundation of China under Grant No.12375143. J. P. V. is supported by the Department of Energy under Grant No. DE-SC0023692. A major portion of the computational resources were also provided by Sugon Advanced Computing Center.

\biboptions{sort&compress}
\bibliographystyle{elsarticle-num}
\bibliography{ProtonGPDletter.bib}
\end{document}